\documentclass[aps,pre,twocolumn,showpacs,floatfix,a4paper]{revtex4-1}

\usepackage{amsfonts,amsmath,amssymb}
\usepackage{textcomp}
\usepackage{bm}
\usepackage{upgreek}
\usepackage[dvips]{graphicx}
\usepackage[latin1]{inputenc}

\newcommand{\notop}{{{}_{}}}

\newcommand{\etal}{\textit{et~al.}}

\renewcommand{\vec}[1]{\bm{#1}}
\newcommand{\ee}{\mathrm{e}}
\newcommand{\ii}{\mathrm{i}}
\newcommand{\dm}{\mathrm{d}}
\newcommand{\avr}[1]{\big\langle #1 \big\rangle}

\DeclareMathOperator{\re}{Re}
\DeclareMathOperator{\im}{Im}
\newcommand{\krondel}[2]{\delta^\notop_{#1 #2}}

\newcommand{\iot}{\ii\omega t}

\newcommand{\pp}{\partial^{{}}}
\newcommand{\ppsqr}{\partial^{\,2_{}}}
\newcommand{\ppr}{\partial^\notop_r}
\newcommand{\ppth}{\partial^\notop_\theta}
\newcommand{\nablabf}{\boldsymbol{\nabla}}

\newcommand{\Lapl}{\nabla^2}

\newcommand{\divop}{\nablabf\cdot}
\newcommand{\scap}{\!\cdot\!}

\newcommand{\BBB}{\vec{B}}

\newcommand{\een}{\vec{e}^\notop}
\newcommand{\eer}{\vec{e}^\notop_r}
\newcommand{\eeth}{\vec{e}^\notop_\theta}

\newcommand{\FFFrad}{\vec{F}^\mathrm{rad}}
\newcommand{\Frad}{F^{\mathrm{rad}_{}}}
\newcommand{\Fbuoy}{F^{\mathrm{buoy}_{}}}
\newcommand{\FradID}{F^{\mathrm{rad}_{}}_\mathrm{1D}}

\newcommand{\hI}{h^{1_{}}}

\newcommand{\kkk}{\vec{k}}

\newcommand{\nnn}{\vec{n}}

\newcommand{\pAB}{p^\notop_\textrm{ab}}

\newcommand{\rrr}{\vec{r}}

\newcommand{\uuu}{\vec{u}}

\newcommand{\vvv}{\vec{v}}

\newcommand{\vnn}{{v}^\notop}
\newcommand{\vIn}{v^{\mathrm{in}_{}}}
\newcommand{\vP}{v^\notop_\mathrm{p}}
\newcommand{\VP}{V^\notop_\mathrm{p}}
\newcommand{\vInD}{v^\notop_\mathrm{in}}
\newcommand{\vInsqr}{v^{2_{}}_\mathrm{in}}
\newcommand{\vSc}{v^{\mathrm{sc}_{}}}
\newcommand{\vvvIn}{\vec{v}^\notop_\mathrm{in}}
\newcommand{\vvvInconj}{\vec{v}^{*_{}}_\mathrm{in}}
\newcommand{\vvvSc}{\vec{v}^\notop_\mathrm{sc}}
\newcommand{\vvvP}{\vec{v}^\notop_\mathrm{p}}
\newcommand{\vvvAB}{\vec{v}^\notop_\mathrm{ab}}

\newcommand{\vAB}{v^{\mathrm{ab}_{}}}

\newcommand{\zerovec}{\boldsymbol{0}}
\newcommand{\calD}{\mathcal{D}}

\newcommand{\Eac}{E^{{}}_\mathrm{ac}}
\newcommand{\Eacmin}{E^{\mathrm{min}_{}}_\mathrm{ac}}

\newcommand{\kapP}{\kappa^\notop_\mathrm{p}}
\newcommand{\kapTi}{\tilde{\kappa}}

\newcommand{\Urad}{U^{\mathrm{rad}_{}}}
\newcommand{\Upz}{U^{{}}_\mathrm{pz}}

\newcommand{\deltaTi}{\tilde{\delta}}

\newcommand{\epsO}{\epsilon^\notop_0}

\newcommand{\etaScs}{\eta^{{}}_\mathrm{scs}}
\newcommand{\etaWa}{\eta^{{}}_\mathrm{wa}}
\newcommand{\etaNaCl}{\eta^{{}}_\mathrm{NaCl}}

\newcommand{\sigmaABbf}{\bm{\sigma}^\notop_{\textrm{ab}}}

\newcommand{\sigmaAB}{\tilde{\sigma}^{\textrm{ab}_{}}}

\newcommand{\cO}{c^{{}}_0}
\newcommand{\cOsqr}{c^{\,2_{}}_0}

\newcommand{\cScs}{c^{{}}_\mathrm{scs}}

\newcommand{\fI}{f^{{}}_1}
\newcommand{\fII}{f^{{}}_2}
\newcommand{\fIIconj}{f^{*_{}}_2}
\newcommand{\fIIr}{f^{\,\mathrm{r}_{}}_2}
\newcommand{\fIIi}{f^{\,\mathrm{i}_{}}_2}

\newcommand{\kapO}{\kappa^{{}}_0}

\newcommand{\pO}{p^{{}}_0}
\newcommand{\pa}{p^\notop_\mathrm{a}}
\newcommand{\pasqr}{p^{2_{}}_\mathrm{a}}

\newcommand{\pI}{p^{{}}_1}
\newcommand{\pIsqr}{p^{\,2_{}}_1}
\newcommand{\pIn}{p^\notop_\mathrm{in}}
\newcommand{\pInsqr}{p^{2_{}}_\mathrm{in}}
\newcommand{\pInconj}{p^{*_{}}_\mathrm{in}}

\newcommand{\pII}{p^{{}}_2}

\newcommand{\vI}{v^{{}}_1}
\newcommand{\vIsqr}{v^{\,2_{}}_1}
\newcommand{\vvvI}{\vvv^{{}}_1}

\newcommand{\vvvII}{\vvv^{{}}_2}

\newcommand{\phiI}{\phi^{{}}_1}

\newcommand{\phiIn}{\phi^{{}}_\mathrm{in}}

\newcommand{\phiSc}{\phi^{{}}_\mathrm{sc}}
\newcommand{\phiScsqr}{\phi^{2_{}}_\mathrm{sc}}
\newcommand{\phiMp}{\phi^{{}}_\mathrm{mp}}
\newcommand{\phiDp}{\phi^{{}}_\mathrm{dp}}
\newcommand{\rhoWa}{\rho^\notop_\mathrm{wa}}
\newcommand{\rhoO}{\rho^\notop_0}
\newcommand{\rhoI}{\rho^\notop_1}
\newcommand{\rhoII}{\rho^\notop_2}
\newcommand{\rhoIn}{\rho^\notop_\mathrm{in}}
\newcommand{\rhoSc}{\rho^\notop_\mathrm{sc}}
\newcommand{\rhoP}{\rho^\notop_\mathrm{p}}

\newcommand{\rhoScs}{\rho^\notop_\mathrm{scs}}
\newcommand{\rhoTi}{\tilde{\rho}}

\newcommand{\SICel}{^\circ\!\textrm{C}}

\newcommand{\SIkg}{\textrm{kg}}

\newcommand{\SIm}{\textrm{m}}

\newcommand{\SImum}{\textrm{\textmu{}m}}

\newcommand{\SImPas}{\textrm{mPa}\:\textrm{s}}

\newcommand{\SIs}{\textrm{s}}

\newcommand{\beq}[1]{\begin{equation} \eqlab{#1}}
\newcommand{\eeq}{\end{equation}}
\newcommand{\bsub}{\begin{subequations}}
\newcommand{\esub}{\end{subequations}}
\def\bal#1\eal{\begin{align}#1\end{align}}
\def\bsubal#1\esubal{\bsub \begin{align}#1\end{align} \esub}
\newcommand{\nn}{\nonumber}
\newcommand{\eqlab}[1]{\label{eq:#1}}
\renewcommand{\eqref}[1]{Eq.~(\ref{eq:#1})}
\newcommand{\eqsref}[2]{Eqs.~(\ref{eq:#1}) and~(\ref{eq:#2})}

\newcommand{\figref}[1]{Fig.~\ref{fig:#1}}

\newcommand{\figlab}[1]{\label{fig:#1}}

\newcommand{\seclab}[1]{\label{sec:#1}}
\newcommand{\tabref}[1]{Table~\ref{tab:#1}}

\newcommand{\tablab}[1]{\label{tab:#1}}

\begin{document}
\title{On the forces acting on a small particle
in an acoustical field in a viscous fluid}

\author{Mikkel Settnes}
\author{Henrik Bruus}
\affiliation{Department of Micro- and Nanotechnology,
Technical University of Denmark,
DTU Nanotech Building 345 B,
DK-2800 Kongens Lyngby, Denmark}

\date{27 October 2011}

\begin{abstract}
We calculate the acoustic radiation force from an ultrasound wave on a compressible, spherical particle suspended in a viscous fluid. Using Prandtl--Schlichting boundary-layer theory, we include the kinematic viscosity of the solvent and derive an analytical expression for the resulting radiation force, which is valid for any particle radius and boundary-layer thickness provided that both of these length scales are much smaller than the wavelength of the ultrasound wave (mm in water at MHz frequencies). The acoustophoretic response of suspended microparticles is predicted and analyzed using parameter values typically employed in microchannel acoustophoresis.
\end{abstract}
\maketitle

\section{Introduction}
Particles suspended in acoustic fields are subject to time-averaged forces from scattering of the acoustic waves. Theoretical studies of these forces, known as acoustic radiation forces, date back to King in 1934, who considered incompressible particles suspended freely in an inviscid fluid \cite{King1934}. In 1955 Yosioka and Kawasima extended the analysis to include the compressibility of the suspended particles \cite{Yosioka1955}. These results were summarized and generalized by a simple and physically intuitive method by Gorkov in 1962 \cite{Gorkov1962}, but limited to inviscid fluids and particles  smaller than the acoustic wavelength $\lambda$.

With recent developments in microfabrication technologies allowing for integration of ultrasound resonators in lab-on-a-chip systems, the acoustic radiation force has received renewed attention as a label- and contact-free way to manipulate particles. Several biotechnological applications of particle trapping and separation have been reported where ultrasound resonances in microchannels were used to create acoustic fields giving rise to acoustic radiation forces on suspended particles. Examples are on-chip acoustophoretic cell separation devices \cite{Petersson2004, Petersson2005}, cell trapping \cite{Hultstrom2007, Evander2007, Svennebring2009}, plasmapheresis \cite{Lenshof2009}, forensic analysis \cite{Norris2009}, food analysis \cite{Grenvall2009}, cell sorting using surface acoustic waves \cite{Franke2010}, cell synchronization \cite{Thevoz2010}, and cell differentiation \cite{Augustsson2010}. At the same time, substantial advancements in understanding the fundamental physics of biochip acoustophoresis have been achieved through full-chip imaging of acoustic resonances \cite{Hagsater2007},  surface acoustic wave generation of standing waves \cite{shi2009}, multi-resonance chips \cite{Manneberg2009}, advanced frequency control \cite{Manneberg2009a, Glynne-Jones2010}, on-chip integration with magnetic separators \cite{Adams2009}, acoustics-assisted microgrippers \cite{Oberti2010}, \emph{in-situ} force calibration \cite{Barnkob2010}, and automated micro-PIV systems \cite{Augustsson2011}.

Traditionally, the acoustic radiation force has been modeled using the inviscid theory of the acoustic radiation force. This approach is approximately correct for particles significantly larger than the thickness $\delta$ of the acoustic boundary layer, in which viscosity do play a dominant role. For a fluid with kinematic viscosity (or momentum diffusivity) $\nu$ and with an acoustic field of angular frequency $\omega$, the boundary layer thickness (or viscous penetration depth) is the momentum diffusion length given by \cite{LordRayleigh1884,Landau1993} 
 \beq{delta_def}
 \delta = \sqrt{\frac{2\nu}{\omega}} \approx 0.6~\SImum,
 \eeq
where the value is for 1-MHz ultrasound in water at room temperature. It is therefore expected that particles or cells with a radius larger than 3~$\SImum$ can be described fairly accurately by the inviscid theory. However, as the technological development pushes for higher accuracy, more refined applications, and the handling of smaller particles, it becomes relevant to calculate the effects on acoustophoresis from viscosity of the solvent.

We are not the first to analyze how  the acoustic radiation force depends on $\delta$. However, the earlier works by Doinikov \cite{Doinikov1997} and by Danilov and Mironov  \cite{Danilov2000} focus mainly on developing general theoretical schemes for particles of radius $a$ smaller than the wavelength $\lambda$ and only provide analytical expressions in the special limits $\delta \ll a \ll \lambda$ and $ a \ll \delta \ll \lambda$. However, given the length scale above, the range of applicability of the published expressions for viscous corrections is \textit{a priori} severely limited. The aim of this paper is to provide an analytical expression for the viscous corrections to the acoustic radiation force on small suspended particles $\delta, a \ll \lambda$, and to analyze its implications for experimentally relevant parameters central for current studies in the field of microchannel acoustophoresis of compressible particles in liquids.

We begin by establishing the governing equations for acoustophoresis in the framework of second-order perturbation theory of the Navier--Stokes equation in the acoustic field. Then, following the inviscid analysis by Gorkov \cite{Gorkov1962}, we express the radiation force on a particle in terms of the far-field solution of inviscid acoustic wave scattering theory, extend this solution to the near-field region close to the particle, and match it with the solution to the incompressible viscous flow problem in the acoustic boundary layer of the particle. From this we obtain the analytical expression for the acoustic radiation force in the viscous case. Finally, we analyze the predictions of the theory for experimentally relevant parameter values.

\section{Perturbation expansion of the governing equations}
\seclab{PertTheory}

The formulation of the governing equations for acoustics in perturbation theory is well known, and the reader is referred to the textbooks by Lighthill \cite{Lighthill2002}, Pierce \cite{Pierce1991}, and Landau \& Lifshitz \cite{Landau1993}. Briefly and to establish notation \cite{Bruus2011}, for a given fluid in the absence of external forces and for isothermal conditions, the theory is based on a combination of the thermodynamic equation of state expressing pressure $p$ in terms of density $\rho$, the kinematic continuity equation for $\rho$, and the dynamic Navier--Stokes equation for the velocity field $\vvv$,
 \bsub
 \eqlab{AcoustBasicEq}
 \bal
 \eqlab{AcoustStateEq}
 p &= p(\rho),\\[1mm]
 \eqlab{AcoustContEq}
 \pp_t\rho &= -\nablabf\scap\big(\rho \vvv \big),\\
 \eqlab{AcoustNSEq}
 \rho\pp_t \vvv &=
 -\nablabf p -\rho(\vvv\scap\nablabf)\vvv
 + \eta\nabla^2 \vvv +
 \beta\eta\:\nablabf(\nablabf\scap\vvv),
 \eal
 \esub
where $\eta$ is the dynamic viscosity of the fluid and $\beta$ the viscosity ratio typically of the order of unity. Thermal effects are neglected, because the thermal diffusion length in liquids is much smaller than the momentum diffusion length (or viscous penetration depth) $\delta$, see Ref.~\onlinecite{Doinikov1997}.

We consider a quiescent liquid, which before the presence of any acoustic wave has constant density $\rhoO$ and pressure $\pO$. Let an acoustic wave constitute tiny perturbations to first and second order (subscript 1 and 2, respectively) in density $\rho$, pressure $p$, and velocity $\vvv$,
 \bsubal
 \rho & = \rhoO + \rhoI + \rhoII, \\
 p &= \pO + \cOsqr\rhoI + \pII, \\
 \vvv &= \vvvI + \vvvII.
 \esubal
Here, we have introduced the speed of sound $\cO$ of the fluid, the square of which is given by the (isentropic) derivative $\cOsqr = (\pp p/\pp\rho)_s$, ensuring the useful identity
 \beq{p1rho1}
 \pI = \cOsqr\rhoI,
 \eeq
and an explicit expression for the compressibility $\kapO$,
 \beq{kap0def}
 \kapO = -\frac{1}{V}\frac{\pp V}{\pp p} =
 \frac{1}{\rhoO}\frac{\pp \rho}{\pp p} = \frac{1}{\rhoO\cOsqr}.
 \eeq

The first-order perturbation (or linearization) of the continuity and Navier--Stokes equation is,
 \bsub
 \eqlab{Acoust1}
 \bal
 \eqlab{AcoustContEq1}
 \pp_t\rhoI &= -\rhoO\nablabf\scap\vvvI,\\
 \eqlab{AcoustNSEq1}
 \rhoO\pp_t \vvvI &= -\cOsqr\nablabf\rhoI + \eta\nabla^2 \vvvI +
 \beta\eta\:\nablabf(\nablabf\scap\vvvI),
 \eal
 \esub
The first-order acoustic wave equation for $\rhoI$ is obtained by taking the time derivative $\pp_t$ of \eqref{AcoustContEq1} followed by insertion of \eqref{AcoustNSEq1} in the resulting expression,
 \beq{WaveEqRhoI}
 \ppsqr_t\rhoI = \cOsqr\bigg[1+\frac{(1+\beta)\eta}{\rhoO\cOsqr}
 \pp_t\bigg]\Lapl\rhoI.
 \eeq
For acoustics fields in the bulk, several times $\delta$ from rigid boundaries, the viscous dissipation is negligible because of the minute damping coefficient, $\eta \omega/(\rhoO\cOsqr) \ll 1$, where $\omega$ is a characteristic angular frequency of the system. For distances within a few times $\delta$ of a rigid wall, the no-slip boundary condition forces the velocity of the fluid to equal that of the wall, and large velocity gradients may occur in \eqref{AcoustNSEq1}, such that  $\eta \vI/\delta^2 \gtrsim \rhoO \pp_t\vI$, and viscosity cannot be neglected.

The acoustic radiation force is a time-average effect that does not resolve the oscillatory behavior of the acoustic fields, so in this work we do not need the full second-order perturbation of the governing equations, but only their time-average. We assume that after the vanishing of transients, any first-order field $f(\rrr,t)$ has a harmonic time dependence,
 \beq{time_harm_def}
 f(\rrr,t) = f(\rrr)\;\ee^{-\iot},
 \eeq
and define the time average $\avr{X}$ over a full oscillation period $\tau$ of a quantity $X(t)$ as
 \beq{avrX}
 \avr{X} \equiv \frac{1}{\tau} \int_0^\tau\dm t\: X(t).
 \eeq
From this we obtain the time-averaged, second-order perturbation of \eqsref{AcoustContEq}{AcoustNSEq} in the form
 \bsub
 \eqlab{AvrAcoustContEq2}
 \bal
 \rhoO \nablabf\scap\avr{\vvvII} &=
 -\nablabf\scap\avr{\rhoI\vvvI},\\[2mm]
 \eqlab{AvrAcoustNSEq2}
 -\nablabf\avr{\pII} + \eta\nabla^2 \avr{\vvvII} &+ \beta\eta\nablabf(\nablabf\scap\avr{\vvvII})
 \nonumber \\
 & = \avr{\rhoI\pp_t\vvvI} +
 \rhoO\avr{(\vvvI\scap\nablabf)\vvvI}.
 \eal
 \esub
We note that the physical, real-valued time average $ \avr{f\:g}$ of two harmonically varying fields $f$ and $g$ with the complex representation \eqref{time_harm_def}, is given by the real-part rule
 \beq{TimeAvrProd}
 \avr{f\:g}
 = \frac{1}{2}\:\re\Big[f(\rrr)\:g^*(\rrr)\Big],
 \eeq
where the asterisk denotes complex conjugation.

Clearly, the time-averaged, second-order fields will in general be non-zero, as the non-vanishing time-averaged products of first-order terms act as source terms on the right-hand side in the governing equations.

In the inviscid bulk, the first-order flow $\vvvI$ is a potential flow, see \eqref{AcoustNSEq1} with $\eta=0$, which when used in \eqref{AvrAcoustNSEq2} and combined with \eqref{kap0def} leads to
 \beq{p2bulk}
 \avr{\pII} = \frac{1}{2}\kapO \avr{\pIsqr}
 - \frac{1}{2}\rhoO \avr{\vIsqr}.
 \eeq

\begin{figure*}[t]
\centering
\includegraphics[]{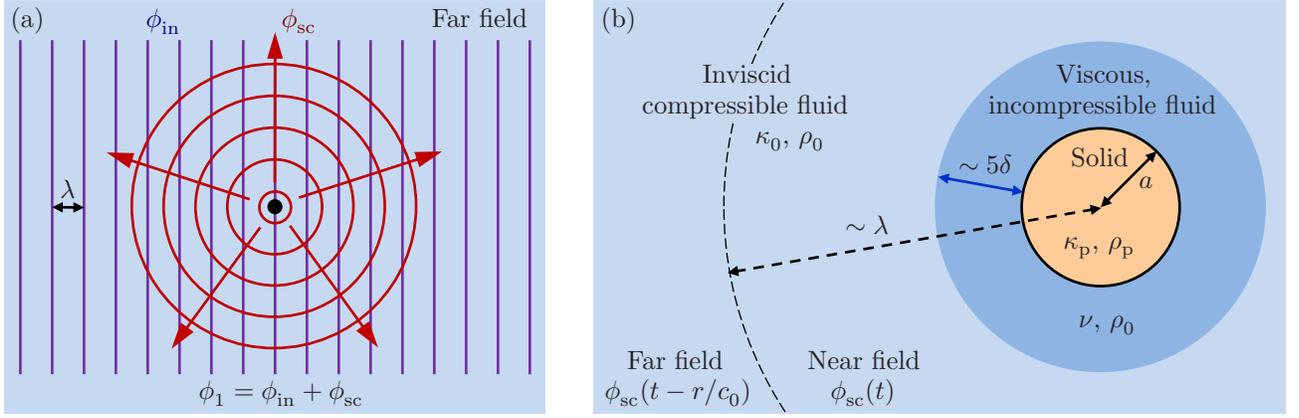}
\caption{\figlab{scattering} (a) Sketch of the far-field region $r \gg \lambda$ of an incoming acoustic wave $\phiIn$ (blue lines) of wavelength $\lambda$ scattering off a small particle (black dot) with radius $a \ll \lambda$, leading to the outgoing scattered wave $\phiSc$ (red circles and arrows). The resulting first-order wave is $\phiI = \phiIn + \phiSc$. (b) Sketch of a compressible spherical particle (yellow) of radius $a$, compressibility $\kapP$, and density $\rhoP$, surrounded by the incompressible, viscous acoustic boundary layer of width $\sim 5\delta$ (dark blue) with density $\rhoO$ and kinematic viscosity $\nu$. Outside is the compressible inviscid bulk (light blue) of compressibility $\kapO$ and density $\rhoO$. The bulk liquid is divided into the near-field region for $r \ll \lambda$, with the instantaneous scattered field $\phiSc(t)$, and the far-field region with time-retarded scattered field $\phiSc(t-r/\cO)$.}
\end{figure*}

\section{The acoustic radiation force}
\seclab{RadForce}

We are analyzing the acoustic radiation force on a compressible, spherical, micrometer-sized particle of radius $a$ suspended in a viscous fluid in an ultrasound field of wavelength $\lambda$ ($=1$~mm in water at room temperature for $\omega/2\pi = 1.5$~MHz), thus $a\ll\lambda$. In terms of acoustic waves, the microparticle thus acts as a weak point scatterer, which we will treat by first-order scattering theory. In response to an incoming wave, described by the oscillatory velocity field $\vvvIn$, an outgoing wave $\vvvSc$ propagates away from the particle. For sufficiently weak waves, the first-order acoustic velocity field $\vvvI$ is given by the sum
 \beq{vIvInvSc}
 \vvvI = \vvvIn + \vvvSc.
 \eeq
Once the first-order scattered field $\vvvSc$ have been determined for a given incoming first-order field $\vvvIn$, the acoustic radiation force  $\FFFrad$ on the particle can be calculated as the time-averaged second-order forces acting on a fixed surface $\pp\Omega$ in the inviscid bulk, encompassing the particle \cite{Gorkov1962}. Momentum conservation and zero bulk forces ensures that any fixed surface can be chosen. For inviscid fluids, $\FFFrad$ is the sum of the time-averaged second-order pressure $\avr{\pII}$ and momentum flux tensor $\rhoO\avr{\vvvI\vvvI}$,
 \bal
 \eqlab{Frad1}
 \FFFrad &=
 -\int_{\pp\Omega}\dm a\: \Big\{
 \avr{\pII}\nnn
 + \rhoO\avr{(\nnn\scap\vvvI)\vvvI}\Big\}
 \\ \nn
 &= -\!\int_{\pp\Omega}\!\!\dm a  \Bigg\{
 \bigg[\frac{\avr{\pIsqr}}{2\rhoO\cOsqr}
 - \frac{\rhoO}{2}\avr{\vIsqr} \bigg]\nnn
 + \rhoO\avr{(\nnn\scap\vvvI)\vvvI}\Bigg\}.
 \eal

To ease the determination of $\pI$ and $\vvvI$, we use that in the inviscid bulk, they can be expressed in terms of a velocity potential $\phiI$ as $\vvvI = \nablabf \phiI$ and $\pI = -\rhoO \pp_t\phiI$. For a harmonic time dependence, \eqref{AcoustNSEq1} implies
 \beq{phiIdef}
 \phiI = -\ii\: \frac{\cOsqr}{\rhoO\omega}\:\rhoI,
 \eeq
and, as sketched in \figref{scattering}(a), we henceforth write
 \bsub
 \bal
 \eqlab{phiI_In_Sc}
 \phiI &= \phiIn + \phiSc,\\
 \eqlab{vI_In_Sc}
 \vvvI &= \nablabf\phiI = \nablabf\phiIn + \nablabf\phiSc,\\
 \eqlab{pI_In_Sc}
 \pI   &= \ii\:\rhoO\omega\:\phiI =
 \ii\:\rhoO\omega\:\phiIn + \ii\:\rhoO\omega\:\phiSc.
 \eal
 \esub

By virtue of \eqsref{WaveEqRhoI}{phiIdef}, $\phiI$ (as well as $\phiIn$ and $\phiSc$) obeys the inviscid wave equation $\ppsqr_t\phi = \cOsqr\Lapl\phi$ in the bulk. As we can use any surface $\pp\Omega$ to calculate the radiation force $\FFFrad$, the simplest choice is a surface in the far-field region $r \gg \lambda$, where the spherical particle of radius $a$ is placed at the center of the coordinate system, and where $\rrr$ is a position vector. According to standard scattering theory, the scattered field $\phiSc$ from a point scatterer can be represented by a time-retarded multipole expansion. In the far-field region, the monopole component and dipole components dominate, $\phiSc \approx \phiMp + \phiDp$, and in general, these two components have the specific forms $\phiMp(\rrr,t) = b(t-r/\cO)/r$ and $\phiDp(\rrr,t) = \nablabf \cdot \big[\BBB(t-r/\cO)/r\big]$, where $b$ is a scalar function and $\BBB$ a vector function of the retarded argument $t-r/\cO$. In first-order scattering theory, $\phiSc$ must be proportional to the first-order fields determined by  $\phiIn$. On physical grounds, the only relevant scalar field is the density, $b \sim \rhoIn$, or equivalently the pressure $\pIn$, while the only relevant vector field is the velocity, $\BBB \sim \vvvIn$.
Here both $\rhoIn$ and $\vvvIn$ are evaluated at the particle position with time-retarded arguments, and in the far-field region $\phiSc$ must therefore have the form
 \bal
 \phiSc(\rrr,t) = & - f_1 \frac{a^3}{3\rhoO}
 \frac{\partial_t \rhoIn(t-r/\cO)}{r} \nonumber \\
 \eqlab{phiScgeneral}
 &-f_2 \frac{a^3}{2}\divop \bigg(\frac{\vvvIn(t-r/\cO)}{r}\bigg), \quad r\gg \lambda,
 \eal
where the particle radius $a$, the unperturbed density $\rhoO$, and the time derivative $\pp_t$ are introduced to ensure the correct physical dimension of $\phiSc$, namely m$^2$/s. The factors $1/3$ and $1/2$ are inserted for later convenience.

Before reaching the main goal of the calculation, namely the determination of the dimensionless scattering coefficients $\fI$ and $\fII$, the radiation force $\FFFrad$ is expressed in terms of the incoming acoustic wave $\phiIn$ at the particle position and the coefficients $\fI$ and $\fII$. When inserting the velocity potentials \eqsref{phiI_In_Sc}{phiScgeneral} into \eqref{Frad1} for $\FFFrad$, we obtain a sum of terms each proportional to the square of $\phiI = \phiIn + \phiSc$. This results in three types of contributions, (\textit{i}) squares of $\phiIn$ containing no information about the scattering and therefore yielding zero, (\textit{ii}) squares of $\phiSc$ proportional to the square of the particle volume $a^6$ and therefore negligible compared to (\textit{iii}) the mixed products $\phiIn\phiSc$ proportional to particle volume $a^3$, which are the dominant contributions to $\FFFrad$. Keeping only these mixed terms representing interference between the incoming and the scattered wave, and using the index notation (including summation of repeated indices), the $i$th component of \eqref{Frad1} becomes
 \bsubal
 \eqlab{Frad21}
 \Frad_i = -\int_{\pp\Omega}&\dm a\: n_j\Bigg\{
 \bigg[\frac{\cOsqr}{\rhoO}\avr{\rhoIn\rhoSc}
 - \rhoO\avr{\vIn_k\vSc_k} \bigg]\krondel{i}{j}
 \nn \\
 &+ \rhoO\avr{\vIn_i\vSc_j} + \rhoO\avr{\vSc_i\vIn_j}\Bigg\}\\
 \eqlab{Frad22}
 = -\int_{\Omega}&\dm \rrr\:\pp_j\Bigg\{
 \bigg[\frac{\cOsqr}{\rhoO}\avr{\rhoIn\rhoSc}
 - \rhoO\avr{\vIn_k\vSc_k} \bigg]\krondel{i}{j}
 \nn \\
 &+ \rhoO\avr{\vIn_i\vSc_j} + \rhoO\avr{\vSc_i\vIn_j}\Bigg\}\\
 \eqlab{Frad23}
 = -\int_{\Omega}&\dm \rrr\: \Bigg\{
 \frac{\cOsqr}{\rhoO}\bigg[
 \avr{\rhoIn\pp_i\rhoSc}+\avr{\rhoSc\pp_i\rhoIn}\bigg]
 \nn \\
 &+\rhoO\bigg[
 \avr{\vIn_i\pp_j\vSc_j} + \avr{\vSc_i\pp_j\vIn_j}\bigg]\Bigg\}\\
 \eqlab{Frad24}
 = -\int_{\Omega}&\dm \rrr\: \Bigg\{
 -\avr{\rhoIn\pp_t\vSc_i} - \avr{\rhoSc\pp_t\vIn_i}
 \nn \\
 &+ \rhoO\avr{\vIn_i\pp_j\vSc_j} - \avr{\vSc_i\pp_t\rhoIn}\Bigg\}\\
 \eqlab{Frad25}
 = -\int_{\Omega}&\dm \rrr\: \Bigg\{
 \avr{\vIn_i\pp_t\rhoSc} + \rhoO\avr{\vIn_i\pp_j\vSc_j}\Bigg\}\\
 \eqlab{Frad26}
 = - \int_{\Omega}&\dm \rrr\: \rhoO\bigg\langle \vIn_i
 \Big(\ppsqr_j\phiSc - \frac{1}{\cOsqr}\ppsqr_t\phiSc
 \Big)\bigg\rangle.
 \esubal
Here, showing details not explained in Ref.~\onlinecite{Gorkov1962}, we have used $\pI = \cOsqr\rhoI$ in \eqref{Frad21}, Gauss's theorem in \eqref{Frad22}, exchange of indices $\pp_iv_k = \pp_i\pp_k\phi = \pp_k\pp_i\phi = \pp_kv_i$ to cancel terms in \eqref{Frad23}, introduction of time derivatives by the continuity equation $\pp_t\rhoI = -\rhoO\pp_j v_{1,j}$ and the Navier--Stokes equation $\rhoO\pp_t v_{1,i} = -\pp_i\pI = -\cOsqr\pp_i\rhoI$ in \eqref{Frad24}, vanishing of time-averages of total time derivatives $\avr{\pp_t(\rho v_i)} = 0$ or $\avr{\rho\pp_t v_i} = -\avr{v_i\pp_t\rho}$ for cancelation and rearrangement in \eqref{Frad25}, and finally reintroduction of the vector potential $\phiSc$ in \eqref{Frad25}.

The d'Alembert wave operator $\ppsqr_j - (1/\cOsqr)\ppsqr_t$ acting on $\phiSc$ appears in the integrand of \eqref{Frad26}, and since $\phiSc$ is a sum of simple monopole and dipole terms, significant simplifications are possible. Just as the Laplace operator acting on the monopole potential $\phi = q/(4\pi\epsO r)$ yields the point-charge distribution, $\ppsqr_j \phi = -(q/\epsO)\delta(\rrr)$, in the static case, the d'Alembert operator acting on the retarded-time monopole and dipole expressions~(\ref{eq:phiScgeneral}) also yields delta function distributions,
 \bal
 \eqlab{dAlembert}
 &\ppsqr_j\phiSc - \frac{1}{\cOsqr}\ppsqr_t\phiSc
 \\ \nn
 & = f_1\:\frac{4\pi a^3}{3\rhoO}\:\pp_t\rhoIn\:\delta(\rrr)
 + f_2\:2\pi a^3\:\nablabf\scap
 \Big[\vvvIn\:\delta(\rrr) \Big],
 \; r\gg\lambda.
 \eal

Now we see the great advantage of working in the far-field limit. The first term is easily integrated, when appearing in \eqref{Frad26}, but for the second term we need to get rid of the divergence operator acting on the delta function before we can evaluate the integral. This we manage by Gauss's theorem. First we note that $\nablabf\cdot\big[v(\rrr)\uuu(\rrr)\big] = v\nablabf\cdot\uuu + \uuu\cdot\nablabf v$ for any scalar function $v$ and vector function $\uuu$. Therefore, $\int_{\pp\Omega}\dm a\:\nnn\cdot(v\uuu) = \int_\Omega\dm\rrr\:\nablabf\cdot(v\uuu) = \int_\Omega\dm\rrr\:(v\nablabf\cdot\uuu+\uuu\cdot\nablabf v)$, and we have derived the expression $\int_\Omega\dm\rrr\:v\nablabf\cdot\uuu = -\int_\Omega\dm\rrr\:\uuu\cdot\nablabf v + \int_{\pp\Omega}\dm a\:\nnn\cdot(v\uuu)$. Now, since $\uuu \propto \vvv\delta(\rrr)$ we obtain in \eqref{Frad26} a volume integral encompassing the delta function, thus yielding a non-zero contribution, and a surface integral avoiding the delta function, thus yielding zero. Consequently, the resulting expression for $\FFFrad$ becomes
 \bsubal
 \eqlab{Frad31}
 \FFFrad &=
 -\frac{4\pi}{3}a^3\:\avr{\fI\vvvIn\pp_t\rhoIn}
 + 2\pi a^3\rhoO\:\avr{\fII(\vvvIn\cdot\nablabf)\vvvIn}\\
 \eqlab{Frad32}
 &=
 \frac{4\pi}{3}a^3\:\avr{\fI\rhoIn\pp_t\vvvIn}
 + 2\pi a^3\rhoO\:\avr{\fII(\vvvIn\cdot\nablabf)\vvvIn}\\
 \eqlab{Frad33}
 &=
 -\frac{4\pi}{3\rhoO\cOsqr}a^3\:\avr{\fI\pIn\nablabf\pIn}
 + 2\pi a^3\:\rhoO\avr{\fII\vvvIn\!\cdot\!\nablabf\vvvIn}\\
 \eqlab{Frad34}
 &=
 -\pi a^3\bigg[
 \frac{2\kapO}{3}
 \re\!\big[\fI\pInconj\nablabf\pIn\big] -
 \rhoO\re\!\big[\fIIconj\vvvInconj
 \!\cdot\!\nablabf\vvvIn\big]\bigg],
 \nn \\
 & \qquad \text{with $\pIn$ and $\vvvIn$ evaluated at}\; \rrr = \zerovec,
 \esubal
where we have integrated over the delta function in \eqref{Frad31}, applied the previously~used rule $\avr{\rhoIn\pp_t\vvvIn} = -\avr{\vvvIn\pp_t\rhoIn}$ in \eqref{Frad32}, inserted $\rhoIn = \pIn/\cOsqr$ and $\pp_t\vvvIn = -\nablabf\pIn/\rhoO$ in \eqref{Frad33}, and finally taken the time average using \eqref{TimeAvrProd} in \eqref{Frad34}.

For standing waves $\phiIn$, the spatial part $f(\rrr)$ of the incoming fields $f(\rrr)\ee^{\iot}$ is real, so the nabla operator in \eqref{Frad34} does not lead to any phase changes. Consequently, the radiation force acting on a small particle ($a\ll\lambda$) placed in a standing wave is a gradient force of the following form,
 \bsub
 \eqlab{FradFinal}
 \begin{align}
 \eqlab{FradGradU}
 \FFFrad &= -\nablabf \Urad, \;
 \text{ for a standing wave $\phiIn$ }\\
 \eqlab{UradFinal}
 \Urad &= \frac{4\pi}{3}a^3 \bigg[
 \re[\fI]\frac{\kapO}{2} \avr{\pInsqr}
 - \re[\fII]\frac{3\rhoO}{4}\avr{\vInsqr}\bigg].
 \end{align}
 \esub
The radiation potential $\Urad$ is proportional to the volume of the particle, and it contains a positive contribution from the acoustic pressure fluctuations and a negative contribution originating from the Bernoulli effect of the acoustic flow speed squared.

For traveling waves, the spatial part $f(\rrr)$ of the incoming fields $f(\rrr)\ee^{\iot}$ contains a phase changing factor, e.g.\ the plane-wave factor $\ee^{\ii\kkk\cdot\rrr}$ or the spherical-wave factor $\ee^{\ii kr}$, which changes the overall structure of the resulting radiation force. Assuming for simplicity an incoming plane wave with wavenumber $\kkk$ parallel to $\vvvIn$, we have $\nablabf \pIn = \ii \kkk \pIn$ and $\nablabf \vvv = \ii \kkk \vvv$, and \eqref{Frad34} leads to a resulting radiation force of the form
 \bal
 \eqlab{FradRun}
 \FFFrad = \frac{4\pi}{3}a^3 & \bigg[
 \im[\fI]\frac{\kapO}{2} \avr{\pInsqr}
 + \im[\fII]\frac{3\rhoO}{4}\avr{\vInsqr}\bigg]\kkk,\\
 & \nn \text{for a purely traveling wave $\phiIn$ }.
 \eal
Note that this is not a gradient force.

\section{The scattering coefficients}
\seclab{f1f2}

The scattering coefficients $\fI$ and $\fII$ of \eqref{phiScgeneral}, are found by matching the pressure $\pI$ and velocity $\vvvI$ of the fluid with the boundary conditions at the particle moving with the instantaneous velocity $\vvvP$. In the following we use a spherical coordinate system with unit vectors $(\eer,\eeth,\een_\phi)$ located at the instantaneous center of the particle. Due to the azimuthal symmetry of the problem, all fields depend only on $r$ and $\theta$, and the velocities has no azimuthal component, $\vvv = \vnn_r\eer + \vnn_\theta\eeth$. The polar axis $\een_z$ points along the instantaneous direction of the incoming velocity $\vvvIn$, such that $\vvvIn = \vInD \een_z$. By the azimuthal symmetry of the problem, the particle must also move in that direction, $\vvvP = \vP \een_z$,
 \bsub
 \begin{alignat}{3}
 \eqlab{vInDir}
 \vvvIn &= \vInD\een_z & &
 = \cos\theta\:\vInD\eer - \sin\theta\:\vInD\eeth,\\
 \eqlab{vPDir}
 \vvvP &= \vP\een_z & &
 = \cos\theta\:\vP\eer - \sin\theta\:\vP\eeth.
 \end{alignat}
 \esub

As sketched in \figref{scattering}(b), the response of the fluid is different in three regions of space \cite{Landau1993}. Just outside the sphere, in the so-called acoustic boundary layer given by $a < r \lesssim a + 5\delta$, viscosity is important due to the increased shear gradients in the velocity fields, as discussed after \eqref{WaveEqRhoI}. Moreover, the fluid appears incompressible, since the time it takes an acoustic wave to propagate across the boundary layer around the particle is much less than the oscillation period, $(a + 5\delta)/\cO \ll 1/\omega$ or $a + 5\delta \ll \lambda$. The first-order pressure and velocity fields in the viscous and incompressible acoustic boundary layer are denoted $\pAB$ and $\vvvAB$, respectively. In the next region, the so-called near-field region with $a + 5\delta \lesssim r \ll \lambda$, the fluid is inviscid and compressible, but $\phiSc$ depends on the instantaneous argument $t$ and not the time-retarded argument $t-r/\cO$. The scattering potential \eqref{phiScgeneral} with its monopole and dipole term becomes
 \bsub
 \eqlab{phiScNear}
 \bal
 \phiSc(r,\theta) &= \phiMp(r) + \phiDp(r,\theta), \quad a+5\delta \lesssim r \ll \lambda,\\
 \eqlab{phiMpNear}
 \phiMp(r) &=
 - f_1 \frac{a^3}{3\rhoO}\partial_t\rhoIn \frac{1}{r},\\
 \eqlab{phiDpNear}
 \phiDp(r,\theta) &=
 + f_2 \frac{a^3}{2}\vInD\frac{\cos\theta}{r^2}.
 \eal
 \esub
Finally, outermost in the far-field region $r \gg \lambda$, the fluid is inviscid and compressible with a time-retarded $\phiSc$.

In first-order scattering theory the monopole and dipole parts of the problem do not mix: $\fI$ is the coefficient in the monopole scattering potential $\phiMp$ from a stationary sphere in the incoming density wave  $\rhoIn$, while $\fII$ is the coefficient in the dipole scattering potential $\phiDp$ from an incompressible sphere moving with velocity $\vvvP$ in the incoming velocity wave $\vvvIn$.

\subsection{The monopole scattering coefficient $\fI$}
\seclab{f1}

The presence of a stationary, compressible particle causes a mass rate $\partial_t m$ of fluid to be ejected, that would otherwise have entered the particle volume. To first order, the ejection is determined by the mass flux $\rhoO\vvvSc$ carried by the scattered wave through a surface $\pp\Omega$ encompassing the particle in the near-field region. For a spherical surface with surface vector $\nnn = \een_r$, we obtain
 \beq{MassFlux_1}
 \pp_t m
 = \int_{\pp\Omega} \dm a\: \een_r \cdot \big(\rhoO\vvvSc\big)
 =  f_1\: \frac{4\pi}{3}a^3\: \pp_t \rhoIn.
 \eeq
The factor $1/3$ was introduced in \eqref{phiScgeneral} to make the particle volume $V_p = (4\pi/3)a^3$ appear here. The rate of ejected mass can also be written in terms of the rate of change of the incoming density $\rho_0 +\rhoIn$ multiplied by the particle volume $V_p$ as $\partial_t m= \partial_t \big[(\rho_0+\rhoIn)V_p\big]$. Expressing this through the compressibility $\kappa =-(1/V)(\partial V/\partial p)$ of the fluid, $\kapO$, and the particle, $\kapP$, we obtain,
 \beq{MassFlux_2}
 \pp_t m = \bigg[1-\frac{\kapP}{\kapO}\bigg]
 V_p\: \partial_t \rhoIn.
 \eeq
Now, $\fI$ is obtained by equating \eqsref{MassFlux_1}{MassFlux_2},
 \beq{f1}
 \fI(\kapTi) =  1 - \kapTi, \quad \text{with }
 \kapTi = \frac{\kapP}{\kapO}.
 \eeq
This result is identical to that of Gorkov \cite{Gorkov1962}.
We note that $\fI$ is real-valued and depends only on the compressibility ratio $\kapTi$ between the particle and the fluid; the viscosity of the fluid does not influence the compressibility and mass ejection. For identical compressibilities, $\kapTi = 1$, the monopole scattering vanishes, $\fI(1) = 0$.

\subsection{The dipole scattering coefficient $\fII$}
\seclab{f2}

As $\fII$ is related to the translational motion of the particle, it depends on the viscosity of the fluid, and we must therefore explicitly calculate the first-order velocity $\vvvAB(r,\theta)$ in the viscous, acoustic boundary layer $a < r \lesssim a + 5\delta$. This velocity must match the dipole part $\vvvIn + \nablabf \phiDp$ of the fluid velocity in the near-field region $r \ll \lambda$, see \eqsref{vInDir}{phiDpNear}. Because of the separation of length scales the matching can be made at a distance $r \approx r^*$ fulfilling $a + 5\delta \ll r^* \ll \lambda$, and this so-called asymptotic matching \cite{VanDyke1975, Landau1993} is written
 \bsub
 \eqlab{asympMatch}
 \bal
 \eqlab{asympMatchR}
 \eer\cdot \vvvAB(r \approx r^*,\theta) &=
 \bigg[1 - \fII\:\frac{a^3}{r^3} \bigg]\:\vInD\cos\theta,\\
 \eqlab{asympMatchTH}
 \eeth\cdot \vvvAB(r \approx r^*,\theta) &=
 \bigg[1 + \frac{1}{2}\fII\:\frac{a^3}{r^3} \bigg]\vInD\:(-\sin\theta).
 \eal
 \esub
At the surface of the sphere, $r=a$,  the no-slip boundary condition requires $\vvvAB$ to equal $\vvvP$ of \eqref{vPDir},
 \bsub
 \eqlab{sphereMatch}
 \bal
 \eqlab{sphereMatchR}
 \eer\cdot \vvvAB(a,\theta) &= \vP\cos\theta,\\
 \eqlab{sphereMatchTH}
 \eeth\cdot \vvvAB(a,\theta) &= \vP(-\sin\theta).
 \eal
 \esub
The velocity $\vP$ of the sphere is given by Newton's second law with $\pp_t\vP = -\ii\omega \vP$ and with the viscous stress from the fluid acting on the surface of the sphere,
 \bal
 \eqlab{Newton2}
 & -\ii\frac{4}{3}\pi\:a^3 \rhoP \omega\:\vP =
 \int_{\pp\VP}\dm a\: \nnn \cdot \sigmaABbf\cdot\een_z
 \\ \nonumber
 & \quad =
 2\pi\:a^2 \int_{-1}^1 \dm(\cos\theta)
 \Big[\big(-\pAB+\sigmaAB_{rr}\big)\cos\theta
 -\sigmaAB_{\theta r}\sin\theta\Big].
 \eal
The viscous stress components are $\sigmaAB_{rr}=2\eta \pp_r \vAB_r$ and $\sigmaAB_{\theta r} = \eta \big[ (1/r)\pp_\theta \vAB_r + \partial_r \vAB_\theta-(1/r) \vAB_\theta\big]$.

The determination of the pressure $\pAB$ and velocity field $\vvvAB$ is eased by the incompressibility of the fluid in the acoustic boundary layer, $\nablabf\cdot\vvvAB = 0$. Taking the divergence of the first-order Navier--Stokes equation~(\ref{eq:AcoustNSEq1}), leads to the Laplace equation of the pressure, $\Lapl\pAB = 0$. Since we are seeking the dipole solution, and since $\pAB$ must match asymptotically with the dipole part $\ii\rhoO\omega \big(\phiIn+\phiDp)$ of \eqsref{pI_In_Sc}{phiScNear}, we can immediately write down the pressure in the boundary layer,
 \bsub
 \bal
 \eqlab{pABr}
 \pAB(r,\theta) &=
 \ii\rhoO\omega \Big[r+\frac{1}{2}\frac{a^3}{r^2}\fII\Big]\vInD\cos\theta,\\
 \eqlab{pAB}
 \pAB(a,\theta) &= \ii\rhoO\omega a\Big[1+\frac{1}{2}\fII\Big]\vInD\cos\theta.
 \eal
 \esub

For the velocity field, incompressibility combined with the azimuthal symmetry of the problem, implies that $\vvvAB$ can be written in terms of a stream function $\Psi(r,\theta)$ as
 \bsubal
 \eqlab{Stream}
 \vvvAB(r,\theta) &= \nablabf \times
 \big[\Psi(r,\theta)\: \een_\phi \big]
 \\
 \eqlab{StreamComp}
 &= \frac{\ppth(\sin\theta\:\Psi)}{r\sin\theta}\:\eer
 - \frac{\ppr(r\:\Psi)}{r}\:\eeth
 \esubal
Taking the rotation of the first-order Navier--Stokes equation~(\ref{eq:AcoustNSEq1}) and using \eqref{Stream} leads to an equation for $\Psi$,
 \beq{NSvAB}
 \Lapl \big(\Lapl +q^2\big) \big[\Psi(r,\theta)\: \een_\phi \big] = \zerovec,
 \;\text{with }\; q = \frac{1+\ii}{\delta}.
 \eeq
The solution to this biharmonic vector equation can be found as the sum $\Psi = \Psi^{{}}_1 + \Psi^{{}}_2$, where
 \bsub
 \eqlab{PsiEqns}
 \begin{alignat}{3}
 \eqlab{Psi1eqn}
 \Lapl (\Psi^{{}}_1\een_\phi ) & = 0
 & &\;\Rightarrow\; &
 \Lapl\Psi^{{}}_1 - \frac{\Psi^{{}}_1}{r^2\sin^2\theta} &= 0,\\
 \eqlab{Psi2eqn}
 \big(\Lapl +q^2\big) (\Psi^{{}}_2 \een_\phi ) &=0
 & &\;\Rightarrow\; &
 \Lapl\Psi^{{}}_2 - \frac{\Psi^{{}}_2}{r^2\sin^2\theta} &=
 -q^2\Psi^{{}}_2.
 \end{alignat}
 \esub
The dipole part of the solution to \eqref{Psi1eqn} is the Legendre form $\Psi^{{}}_1(r,\theta) = A^\notop_1r\cos\theta + A^\notop_2 \cos\theta/r^2$. Likewise, the dipole part of the solution to \eqref{Psi2eqn} is the Hankel form $\Psi^{{}}_2(r,\theta) = B \hI_1(qr)\: a \vInD\sin\theta$, where $\hI_1(s) = -\ee^{\ii s}(s+\ii)/s^2$ is the spherical Hankel function of the first kind (outgoing wave) of order 1. The constants $A^\notop_1$ and $A^\notop_2$ are given by the asymptotic matching conditions (\ref{eq:asympMatch}), so $\Psi$ and $\vvvAB = \nablabf\times(\Psi\een_\phi)$ become
 \bsub
 \eqlab{PsivAB}
 \bal
 \Psi(r,\theta) &=
 \bigg[\frac{1}{2}r - \frac{\fII}{2}\frac{a^3}{r^2}
 + a\hI_1(qr)\:B \bigg]\:\vInD\sin\theta,\\
 \eqlab{vABr}
 \vvvAB\!\cdot\!\een_r &=
 \bigg[1 - \fII\frac{a^3}{r^3} +
 2qaB\bigg(\frac{\hI_1(s)}{s}\bigg)^{{}}_{\!qr}\bigg]\vInD\cos\theta,\\
 \eqlab{vABth}
 \vvvAB\!\cdot\!\een_\theta &=
 \bigg[1 + \frac{\fII}{2}\frac{a^3}{r^3} +
 qaB\bigg\{\!
 \frac{\pp_s\big(s\hI_1(s)\big)}{s}
 \!\bigg\}^{{}}_{\!qr}\bigg]
 \vInD(-\sin\theta).
 \eal
 \esub
Note that all information about the viscous boundary layer is given through the constant $q = (1+\ii)/\delta$ of \eqref{NSvAB}. Since the spherical Hankel function decays exponentially on the length scale $\delta$, $\hI_1(qr) \propto \ee^{\ii qr} \propto \ee^{-qr}$, the viscous, acoustic boundary layer has a width of $\sim 5\delta$. An important, dimensionless parameter is the ratio $\deltaTi$ of the viscous penetration length $\delta$ and the particle radius~$a$,
 \beq{deltaTi}
 \deltaTi = \frac{\delta}{a}.
 \eeq

Inserting $\vvvAB$ of \eqsref{vABr}{vABth} into the no-slip condition at the surface of the sphere, \eqref{sphereMatch}, and the pressure $\pAB$ of \eqref{pAB} together with $\vvvAB$ into Newton's second law for the sphere, \eqref{Newton2}, we arrive at three equations for the three unknowns $\fII$, $\vP$, and $B$,
 \bsub
 \eqlab{BvPf2}
 \bal
 \eqlab{vP1}
 \vP &= \bigg[1-\fII+2B\hI_1(qa) \bigg]
 \:\vInD,\\
 \eqlab{vP2}
 \vP &= \bigg[1+\frac{1}{2}\fII+
 B\pp_s\big(s\hI_1(s)\big)^{{}}_{qa}\bigg]
 \:\vInD,\\
 \eqlab{vPN2}
 \rhoTi\vP &= \bigg[1+\frac{1}{2}\fII+
 \ii\deltaTi^2 G(qa)\:B\bigg]
 \:\vInD,
 \eal
 \esub
where the auxiliary function $G(s)$ is given by
 \bal
 G(s) &= 2s^2\pp_s\bigg[\frac{\hI_1}{s}\bigg] + 2\hI_1
 + s^2\pp_s\bigg[\frac{\pp_s(s\hI_1)}{s}\bigg] - \pp_s(s\hI_1)
 \nonumber\\
 &= s\ppsqr_s(s\hI_1) -2\hI_1 = \ii s (\ii+s)\: \hI_0(s),
 \eal
with $\hI_0(s) = -(\ii/s)\ee^{\ii s}$. When subtracting \eqref{vP1} from \eqref{vP2}, $\pp_s\big(s\hI_1(s)\big) -2\hI_1(s) = s\hI_0(s) - 3\hI_1(s)$ appears. It is therefore useful to introduce the dimensionless, $\deltaTi$-dependent variable $\gamma$ given by
 \beq{gammaDef}
 \gamma(\deltaTi) = \frac{3\hI_1(qa)}{qa\hI_0(qa)}
 = -\frac{3}{2} \Big[1+\ii(1+\deltaTi)\Big] \deltaTi.
 \eeq
By straightforward algebra, $\fII$ is found from \eqref{BvPf2},
 \beq{f2Final}
 \fII(\rhoTi,\deltaTi) = \frac{2\big[1-\gamma(\deltaTi)\big](\rhoTi-1)}{2\rhoTi+1-3\gamma(\deltaTi)}.
 \eeq
The viscosity-dependent dipole scattering coefficient $\fII$ is in general a complex-valued number, and its real and imaginary values  are abbreviated as
 \bsubal
 \eqlab{f2rdef}
 \fIIr(\rhoTi,\deltaTi) &= \re\big[\fII(\rhoTi,\deltaTi)\big],\\
 \eqlab{f2idef}
 \fIIi(\rhoTi,\deltaTi) &= \im\big[\fII(\rhoTi,\deltaTi)\big].
 \esubal
In the absence of viscosity, $\deltaTi = 0$, we recover the real-valued result by Gorkov \cite{Gorkov1962},
 \beq{f2Gorkov}
 \fII(\rhoTi,0) =
 \frac{2(\rhoTi-1)}{2\rhoTi+1}.
 \eeq
Physically, this result for an inviscid fluid can also be derived directly from \eqref{BvPf2}: The acoustic boundary layer vanishes, $B = 0$, and the condition~(\ref{eq:vP2}) on the tangential velocity component is dropped, so we are left with the normal-component condition~(\ref{eq:vP1}), $\vP = (1-\fII)\vInD$, and Newton's second law~(\ref{eq:vPN2}), $\rhoTi\vP = \big(1+\frac{1}{2}\fII\big)\vInD$, from which \eqref{f2Gorkov} follows.

The non-zero imaginary part $\fIIi$ of $\fII$ for $\deltaTi > 0$ implies that for a traveling wave $\phiIn$ the product terms $\phiIn\phiSc$ of order $a^3$ remain finite. This is in contrast to the inviscid case, where these terms vanish and the radiation force is reduced by a factor of $(ka)^3 = (2\pi a/\lambda)^3$ since only the quadratic terms $\phiScsqr$ proportional to $a^6$ remain finite \cite{Yosioka1955, Gorkov1962}. In agreement with Doinikov \cite{Doinikov1997}, our analysis thus predicts the possibility of realizing a radiation force for traveling waves, which is a factor of $(ka)^{-3} = (\lambda/2\pi a)^3$ stronger than expected from the standard inviscid theory.

\subsection{Properties of $\bm{\fII}$}

As the dipole scattering coefficient $\fII$, in contrast to the monopole coefficient $\fI$, depends on viscosity, we study some of its properties in more detail below, see \figref{f2_vs_delta}. Importantly, $\fII$ is zero for neutral-buoyancy particles ($\rhoTi = 1$) irrespective of the viscosity,
 \beq{NeutralBuoyancy}
 \fII(1,\deltaTi) = 0,
 \eeq
and generally small for near-neutral-buoyancy particles.

For a large particle in a low-viscosity fluid $\deltaTi \ll 1$, the correction to the Gorkov expression for $\fII$ is found by Taylor-expanding \eqref{f2Final} to first order in $\deltaTi$,
 \beq{f2_low_visc}
 f_2(\rhoTi,\deltaTi\ll 1) \approx \frac{2(\rhoTi-1)}{2\rhoTi+1}
 \bigg[1+\frac{3(\rhoTi-1)}{2\rhoTi+1}(1+\ii)\:\deltaTi\bigg],
 \eeq
in agreement with Doinikov \cite{Doinikov1997}. In earlier work by Weiser and Apfel \cite{Weiser1982} (building on Urick \cite{Urick1948}) the numerator of the viscous correction was found to be $(9/2)\deltaTi$ instead of $3(\rhoTi-1)\deltaTi$. However, this discrepancy is due to an imprecise treatment of the viscous boundary layer in the earlier work.

\begin{figure}[t]
\centering
\includegraphics[]{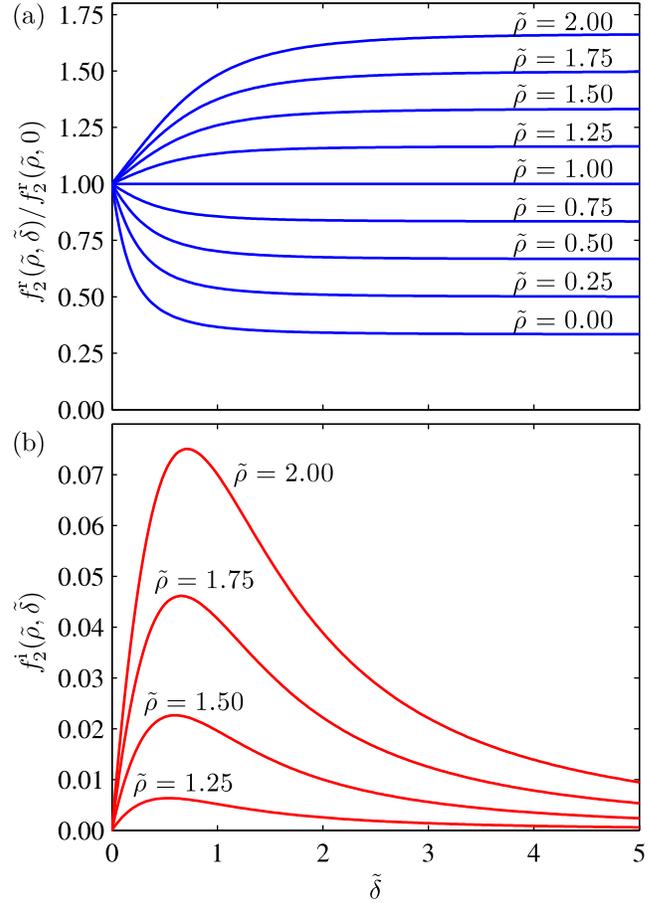}
\caption{\figlab{f2_vs_delta} (a) The real part of the viscosity-dependent dipole scattering coefficient relative to its inviscid counterpart $\fIIr(\rhoTi,\deltaTi)/\fIIr(\rhoTi,0)$ plotted versus the non-dimensionalized thickness $\deltaTi$ of the viscous, acoustic boundary layer. (b) The imaginary part $\fIIi(\rhoTi,\deltaTi)$ versus $\deltaTi$ for $\rhoTi = 1.25$, 1.50, 1.75, and 2.00. For the same parameters $0.17 < \fIIr(\rho,\infty) < 0.67$.}
\end{figure}

For a small particle in a high-viscosity fluid $\deltaTi \gg 1$, the viscosity dependence saturates, as the particle essentially becomes a point singularity at the center of the acoustic boundary, and $\fII$ becomes
 \beq{f2_high_visc}
 f_2(\rhoTi,\deltaTi\gg 1) \approx \frac{2}{3}\:(\rhoTi - 1).
 \eeq
Doinikov presented the same expression except for the overall sign.  This may be a misprint in his paper, as the consequence would otherwise be an un-physical reversal of the sign of the force as $\deltaTi$ is increased from zero to infinity.

In \figref{f2_vs_delta}(a) is shown plots of the real part $\fIIr(\rhoTi,\deltaTi)$ of the viscosity-dependent dipole scattering coefficient relative to the inviscid coefficient $\fIIr(\rhoTi,0)$ as a function of $\deltaTi$ for different values of $\rhoTi$ between 0 and 2. For these values of $\rhoTi$ the values of $\fIIr(\rhoTi,\deltaTi)/\fIIr(\rhoTi,0)$ fall in the range from 0.3 to 1.7, and the saturation of $\fIIr$ sets in for moderate values of $\deltaTi$ between 1 and 2.

In \figref{f2_vs_delta}(b) is shown the imaginary part $\fIIi(\rhoTi,\deltaTi)$,
 \beq{fIIiFull}
 \fIIi(\rhoTi,\deltaTi) = \frac{6(1-\rhoTi)^2(1+\deltaTi)\deltaTi}{
 (1+2\rhoTi)^2+9(1+2\rhoTi)\deltaTi+\frac{81}{2}(\deltaTi^2+
 \deltaTi^3+\frac{1}{2}\deltaTi^4)},
 \eeq
with
 \bsubal
 \fIIi(\rhoTi,\deltaTi \ll 1) &\approx
 \frac{6(1-\rhoTi)^2}{(1+2\rhoTi)^2}\:\deltaTi,\\
 \fIIi(\rhoTi,\deltaTi \gg 1) &\approx
 \frac{24}{81}(1-\rhoTi)^2\:\deltaTi^{-2}.
 \esubal
It exhibits a marked maximum for $\deltaTi \approx 0.5$ with an amplitude roughly one order of magnitude smaller than the saturation value of $\fIIr(\rhoTi,\infty)$, \eqref{f2_high_visc}, for the corresponding densities $\rhoTi$.

\subsection{Resulting expressions for the radiation force}

In summary, the main result of the paper are the following analytical expressions for the acoustic radiation force $\FFFrad$ on a spherical particle of radius $a$, density $\rhoP$, and compressibility $\kapP$ suspended in a fluid of density $\rhoO$, compressibility $\kapO$, and viscosity $\eta$, and exposed to a first-order standing and traveling acoustic wave $\pIn$ and $\vvvIn$ in the long wavelength limit $\lambda \gg a$.

For a standing acoustic wave we have obtained
 \bsub
 \eqlab{FradRES}
 \begin{align}
 \eqlab{FradGradURes}
 \FFFrad &= -\nablabf \Urad,\\
 \eqlab{UradRES}
 \Urad &= \frac{4\pi}{3}\:a^3 \bigg[
 \fI\:\frac{1}{2}\kapO\: \avr{\pInsqr}
 - \fIIr\:\frac{3}{4}\rhoO\:\avr{\vInsqr}\bigg],\\
 \eqlab{f1RES}
 \fI(\kapTi) &=  1 - \kapTi,
 \text{ with }
 \kapTi = \frac{\kapP}{\kapO},\\
 \eqlab{f2Res}
 \fIIr(\rhoTi,\deltaTi) &= \re\!\Bigg[\!\frac{2\big[1\!-\!\gamma(\deltaTi)\big]
 (\rhoTi\!-\!1)}{2\rhoTi+1-3\gamma(\deltaTi)}\Bigg],
 \text{ with }
 \rhoTi = \frac{\rhoP}{\rhoO},\\
 \eqlab{gamRES}
 \gamma(\deltaTi) &=
 -\frac{3}{2} \Big[1+\ii(1+\deltaTi)\Big] \deltaTi,
 \text{ with }
 \deltaTi = \frac{\delta}{a},
 \end{align}
 \esub
and for a traveling planar wave with wavevector $\kkk$,
 \beq{FradTravel}
 \FFFrad = \fIIi(\rhoTi,\deltaTi)\: \pi a^3 \rhoO \avr{\vInsqr}\kkk,
 \eeq
where $\fIIi(\rhoTi,\deltaTi)$ is given by \eqref{fIIiFull}.

\section{Experimental implications}
\label{sec:exp}

\subsection{Typical materials}

In practical applications, especially involving biological samples, the solvents are often aqueous salt solutions. Among these, sodium chloride (NaCl) solutions are arguably the ones best characterized acoustically, so we use this solvent as one of our model liquids in the following analysis. In Appendix~\ref{Sec:NaCl_solutions} the current best values of the speed of sound $\cScs$, the density $\rhoScs$, and the viscosity $\etaScs$ of  sodium chloride solutions (scs) are given as a function of temperature $T$ (in $\SICel$) and mass fraction $S$ of NaCl in the solution (salinity). To study the effects of change in viscosity we also include calculations with glycerol (gl) and with percoll (pc), a solution of polyvinylpyrrolidone-coated silica nanoparticles for which the speed of sound is nearly the same as for pure water \cite{Laurent1980}, see \tabref{parameters}.

\begin{table}[t]
\caption{\tablab{parameters} List of values of material parameters at $20~\SICel$ for typical liquids [water (wa), NaCl solution (scs), percoll (pc), glycerol (gl)] and solids [pyrex (PY), polystyrene (PS), polymethacrylate (PM), melamine resin (MR), a representative biological cell (Cell)] used in microchannel acoustophoresis.}
\begin{ruledtabular}
\begin{tabular}{ccccc}
Material  &
Density &
Compress. &
Longitud.\ speed &
Viscosity \\
 &
$\rho$ [kg/m$^3$] &
$\kappa$ [1/TPa] &
of sound $c$ [m/s]  &
$\eta$ [mPa$\cdot$s]\\
\hline
wa\footnote[1]{From Ref.~\onlinecite{CRC}}
  & 998.2 & 456  & 1482  & 1.002 \\
scs\footnote[2]{Sodium chloride solution of salinity $S = 0.1$, see Appendix~\ref{Sec:NaCl_solutions}}
  & 1071 & 365  & 1599  & 1.170 \\
pc\footnote[3]{From Sigma-Aldrich Production
   GmbH and Fluka data sheets}
  & 1130 & 390  & 1507  & 100\\
gl\footnotemark[1]
  & 1261  & 219  & 1904  & 1412 \\ \hline
PY\footnotemark[1]
  & 2230  & 27.8 & 5674  & -- \\
PS\footnotemark[3]
  & 1050  & 172  & 2350  & -- \\
PM\footnotemark[3]
  & 1190  & 148  & 2380  & -- \\
MR\footnotemark[3]
  & 1510  & 67.5 & 3132  & -- \\
Cell\footnote[4]{From Refs.~\onlinecite{Godin2007,Augustsson2010}}
  & 1100 & 400  & 1500  & -- \\
\end{tabular}
\end{ruledtabular}
\end{table}

As typical materials for the particles we have chosen to analyze the polymers polystyrene (PS), polymethacrylate (PM), and melamine resin (MR), as well as pyrex (PY) and a typical biological cell (Cell), see \tabref{parameters}.

\subsection{Traveling acoustic waves}

As mentioned above, our theory including viscosity predicts a strong enhancement of the acoustic radiation force by a factor of $(ka)^{-3}$ compared to the standard inviscid theory for the case of purely traveling waves. Here, we analyze the case of a planar traveling wave with $\kkk = k\een_z$ and  $\pI = \pa \ee^{\ii(kz - \omega t)}$. In this case the acoustic energy density $\Eac$ of the wave is $\Eac = \frac{1}{2}\kapO\pasqr$, and the radiation force~(\ref{eq:FradRun}) becomes
 \beq{FradRunZ}
 \FFFrad = \pi a^3 k \fIIi(\rhoTi,\deltaTi)\:\Eac\:\een_z.
 \eeq
For a 5-$\SImum$-diameter pyrex sphere in water at 1.5~MHz we have $a = 2.5~\SImum$, $\deltaTi = 0.23$, $k = 4.24\times10^3$~m$^{-1}$, and $\rhoTi = 2.23$. From \eqref{fIIiFull} we find $\fIIi(2.23,0.23) = 0.058$, and taking the typical acoustic energy density $\Eac = 100$~J/m$^3$ \cite{Barnkob2010}, we arrive at $\Frad \approx 1.2$~pN. Under the influence of this force, the pyrex sphere would reach the terminal translational velocity $\vP = \Frad/(6\pi\eta a) \approx 25~\SImum/\SIs$. This is a significant velocity for microchannel acoustophoresis, where typical velocities lie in the range from 5 to 500~$\SImum$/s \cite{Augustsson2011}. In the standard inviscid theory \cite{Yosioka1955, Gorkov1962}, the estimate for the radiation force is a factor of $(ka)^3 \approx 10^{-6}$ lower, corresponding to an acoustophoretic velocity smaller than 0.1~nm/s.

\subsection{Standing acoustic waves}

In many experiments on rectangular microfluidic channels with coplanar walls at $z=0$ and $z = h$, the incoming wave have approximately been a resonant, standing 1D pressure wave of the form $\pI = \pa\cos(kz)$, with wavenumber $k = n\pi/h$, where $n$ is the number of half wavelengths, and with the acoustic energy density $\Eac = \frac{1}{4}\kapO\pasqr$. The expression for the radiation force then simplifies to the classic result by Yosioka and Kawasima  \cite{Yosioka1955},
 \bsubal
 \eqlab{Frad1D}
 \FradID &= 4\pi\;\Phi(\kapTi,\rhoTi,\deltaTi)\:
 a^3 k\Eac\sin(2kz),\\
 \eqlab{PhiDef}
 \Phi(\kapTi,\rhoTi,\deltaTi) &=
 \frac{1}{3}\:\fI(\kapTi) + \frac{1}{2}\:\fIIr(\rhoTi,\deltaTi),
 \esubal
where the acoustophoretic contrast factor $\Phi(\kapTi,\rhoTi,\deltaTi)$ now depends on viscosity.

Experiments on suspended biological cells involve near-neutral-buoyant particles, $|\rhoTi - 1| \ll 1$, implying that the monopole coefficient $|\fI|$ is typically much larger than the dipole coefficient $|\fIIr|$. Because the acoustic contrast factor $\Phi$ defined in \eqref{PhiDef} is a linear combination of the two scattering coefficients, a good quantitative measure of the ability to detect the effect of viscosity on the acoustic radiation force is therefore the relative change in $\Phi$ with and without viscosity. We therefore find it helpful to introduce the detectability measure $\calD$ of viscous effects as
 \bsubal
 \eqlab{Ddef}
 \calD(\kapTi,\rhoTi,\deltaTi) &=
 \frac{\Phi(\kapTi,\rhoTi,\deltaTi)-\Phi(\kapTi,\rhoTi,0)}{
 \Phi(\kapTi,\rhoTi,0)},\; \text{ or}\\
 \eqlab{Dplus1}
 1 + \calD(\kapTi,\rhoTi,\deltaTi) & = 
 \frac{\Phi(\kapTi,\rhoTi,\deltaTi)}{\Phi(\kapTi,\rhoTi,0)}.
 \esubal
Examples of $\calD$ for in NaCl solutions are shown in \figref{D_vs_conc} going from nearly undetectable sub-1-\%-levels for polystrene spheres to above 10-\%-levels for pyrex spheres.

\begin{figure}[t]
\centering
\includegraphics[]{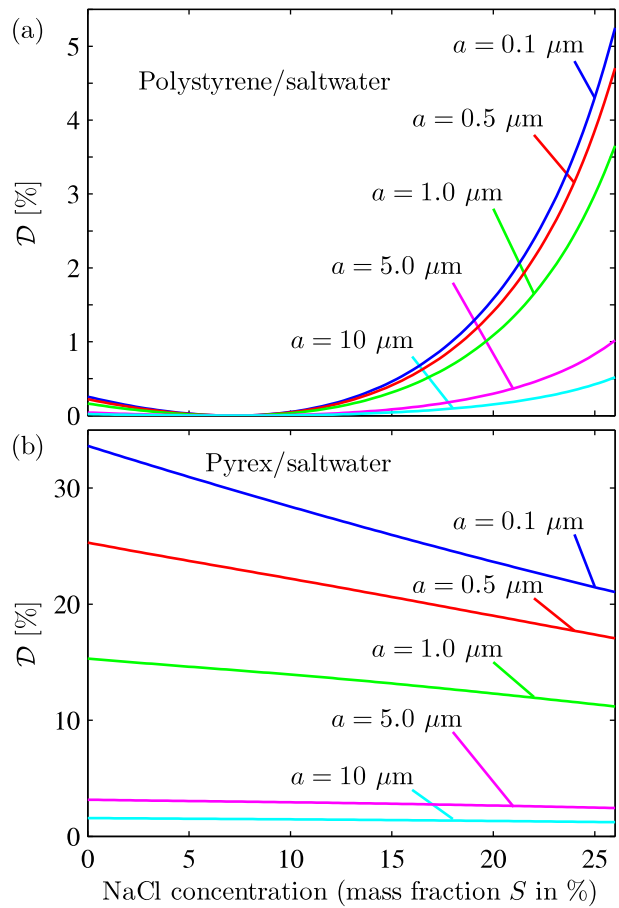}
\caption{\figlab{D_vs_conc} The detectability $\calD$ for solid microspheres of radius $a$ ranging from $0.1~\SImum$ to $10~\SImum$  in a sodium chloride solution as a function of NaCl concentration (mass fraction $S$), see Appendix A. (a) Polystyrene spheres, for which $\calD$ is typically a few percent or lower. (b)  Pyrex spheres, for which $\calD$ easily can be larger than 10\%.}
\end{figure}

\begin{figure}[t]
\centering
\includegraphics[]{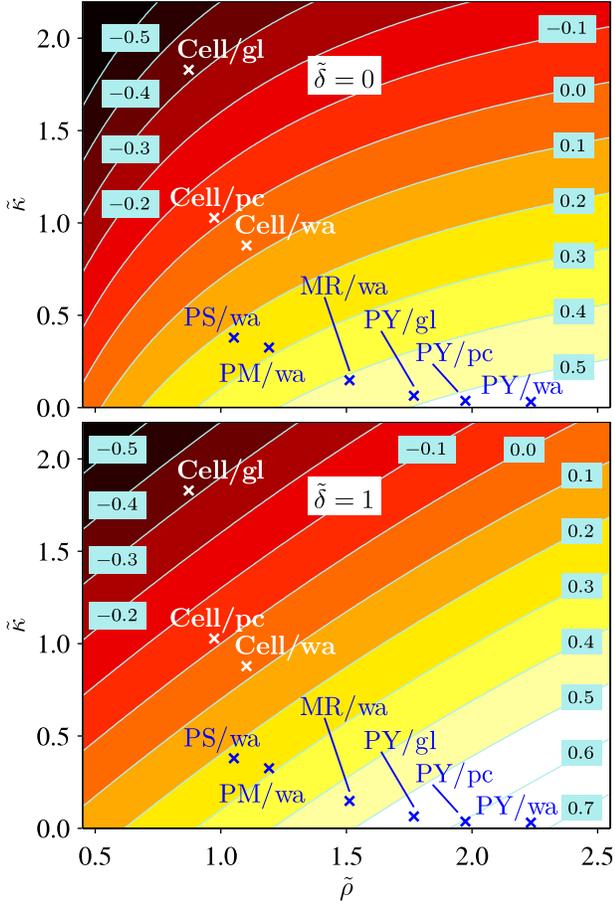}
\caption{\label{fig:PhiPlots} Contour plots of the acoustic contrast factor $\Phi(\kapTi,\rhoTi,\deltaTi)$ as function of $\kapTi$ and $\rhoTi$ for fixed values of $\deltaTi$, from $\Phi < -0.5$ (black) to $\Phi>0.5$ (white) in steps of 0.1. The position in the $(\kapTi,\rho)$-plane for various material parameters are marked by crosses and the following labels: polystyrene (PS), polymethacrylate (PM), melamine resin (MR), a typical biological cell (Cell), water (wa), glycerol (gl), and percoll (pc). (a) The inviscid case $\deltaTi = 0$. (b) The viscous case with $\deltaTi = 1$.}
\end{figure}

The effect of including the viscosity in the expression for the acoustic radiation force can also be illustrated by contour plots of the contrast factor $\Phi$ in the $(\kapTi,\rhoTi)$-plane for fixed values of $\deltaTi$ as shown in \figref{PhiPlots}. The change in the contrast factor is clearly seen by the changing contour lines. While $\Phi$ is independent of $\deltaTi$ along the neutral-buoyancy line $\rhoTi = 1$, its value is increased when going from the inviscid case $\deltaTi = 0$ in panel (a) to the viscous case $\deltaTi = 1$ in panel (b). The change of the contour line $\Phi = 0.0$ is particularly interesting as particles on opposite side of this line move in opposite directions, and the plot of $\Phi$ in the $(\kapTi,\rhoTi)$-plane is therefore also useful when attempting to tune the solvent to obtain binary separation of particles. In \figref{PhiPlots}, a number of specific examples of materials are marked by crosses. As particle material are chosen polystyrene (PS), polymethacrylate (PM), melamine resin (MR), and a typical biological cell (Cell), while the liquids are water (wa), glycerol (gl), and percoll (pc).  Note that the Cell/gl and Cell/wa points lie on opposite sides of the zero contour. A curve connecting these two points would represent the acoustophoretic response of cells in various mixtures of glycerol and water. From a purely physical point of view, this system may therefore form an excellent tunable solvent with respect to obtaining binary separation of cells.

The effect of viscosity can also be studied through vertical acoustic trapping \cite{Apfel1976,Kumar2005}, where the buoyancy force $\Fbuoy = (4\pi/3)(\rhoTi -1)\rhoO a^3 g$ is balanced by a vertically-oriented standing plane-wave acoustic field  $\FFFrad = \Frad\een_z$. For a given acoustic energy density $\Eac$, the maximal acoustic radiation force is given by the amplitude $4\pi \Phi(\kapTi,\rhoTi,\deltaTi) a^3 k \Eac$ in \eqref{Frad1D}. Based on this, the critical trapping force is defined as the threshold for obtaining vertical acoustic trapping using the smallest possible acoustic energy density $\Eacmin$,
 \beq{Eacmin}
 \Eacmin = \frac{|\rhoTi - 1|}{
 3\Phi(\kapTi,\rhoTi,\deltaTi)}\:\frac{\rhoO g}{k}
 \eeq
The effect of viscosity can therefore be measured as,
 \beq{EacminDelta}
 \Eacmin(\deltaTi)
 = \frac{1}{1+\calD}\:\Eacmin(0).
 \eeq
A quantity more readily accessible experimentally may be the voltage $\Upz$ used to drive the piezo transducer generating the ultrasound wave in a typical experiment. As $\Eac$ scales with the square of $\Upz$ \cite{Barnkob2010}, we have\\[-6mm]
 \beq{Uppmin}
 \Upz(\deltaTi)
 = \frac{1}{\sqrt{1+\calD}}\:\Upz(0).
 \eeq
Here, the detectability $\calD$ of \figref{D_vs_conc} appears directly.

\section{Conclusion}

We have derived an analytical expression for the acoustic radiation force in the long-wavelength limit $\delta, a \ll \lambda$ on a compressible, spherical particle of radius $a$ suspended in a liquid with viscous penetration depth $\delta$.

We have analyzed the experimental predictions provided by our expression for traveling waves and for standing waves. In the case of the former we find a strong enhancement proportional to $(ka)^{-3} \approx 10^6$ relative to the inviscid case due to non-vanishing interference between the incoming wave and the scattered wave. For standing waves, we have found a negligible sub-1\% deviation from the inviscid result for large (micrometer-sized), nearly-neutral-buoyancy particles, such as biological cells in water. However, significant deviations above 10\% from the inviscid result were found for buoyant particles, such as pyrex in water. The smaller the particle radius $a$ is relative to the viscous boundary-layer thickness $\delta$, the larger the effect of viscosity.

It should be possible using state-of-the-art instrumentation for microchannel acoustophoresis, such as the automated micro-PIV setup recently published by Augustsson \etal\ \cite{Augustsson2011}, to experimentally test the predictions presented in this paper.

\acknowledgments
We thank Per Augustsson, Lund University, and Rune Barnkob, Technical University of Denmark, for assisting us collecting the parameter values listed in \tabref{parameters}. This work was supported in part by the Danish Council for Independent Research, Technology and Production Sciences, Grant No.~274-09-0342.

\appendix
\section{Material parameters of NaCl solutions}
\label{Sec:NaCl_solutions}

In the following we summarize parameter values for the speed of sound $\cScs$, the density $\rhoScs$, and the viscosity $\etaScs$ of  sodium chloride solutions (NaCl in water) as a function of temperature $T$ (in $\SICel$) and mass fraction $S$ of NaCl in the solution (salinity). The salinity ranges from zero in pure water to a maximum of 0.26 in a saturated solution.

The speed of sound $\cScs(S,T)$ in NaCl/water solutions is given by
Kleis and Sanchez \cite{Kleis1990} as
 \beq{cNaCl}
 \cScs(S,T) = \sum_{j=0}^4 \big(a^{{}}_j + b^{{}}_j S\big)
 \bigg[\frac{T}{1~\SICel}\bigg]^j~\SIm\:\SIs^{-1},
 \eeq
where the coefficients $a^{{}}_j$ and $b^{{}}_j$ are
 \begin{alignat}{4}
 a^{{}}_0 &=\:&  1.40309  &\times10^{3}, &
 b^{{}}_0 &=\:&  1.40190  &\times10^{3}, \\
 a^{{}}_1 &=\:&  4.68391  &\times10^{0}, &
 b^{{}}_1 &=\:& -1.14996 &\times10^{1}, \nn \\
 a^{{}}_2 &=\:& -4.05388 &\times10^{-2}, &
 b^{{}}_2 &=\:&  2.23748 &\times10^{-3}, \nn \\
 a^{{}}_3 &=\:&  1.29550  &\times10^{-5}, &
 b^{{}}_3 &=\:&  1.48238  &\times10^{-3},\nn \\
 a^{{}}_4 &=\:&  6.91485  &\times10^{-7}, \quad&
 b^{{}}_4 &=\:& -9.46165 &\times10^{-6}. \nn
 \end{alignat}

The density $\rhoScs(S,T)$ of sodium chloride solutions is given by Lalibert\'e and Cooper \cite{Laliberte2004} as
 \beq{rhoScs}
 \rhoScs(S,T) = \frac{\rhoWa(T)}{1-S + S\:V^{{}}_\mathrm{app}(S,T)\rhoWa(T)},
 \eeq
where the density $\rhoWa(T)$ of water is given by
 \beq{rhoWa}
 \rhoWa(T) = \frac{1}{1+d\frac{T}{1~\SICel}}
 \sum_{j=0}^5 d^{{}}_j
 \bigg[\frac{T}{1~\SICel}\bigg]^j~\SIkg\: \SIm^{-3},
 \eeq
with the coefficients $d$ and $d^{{}}_j$ being
 \begin{alignat}{4}
 d        &=\:&  1.6879850  &\times10^{-2}, & &&& \\
 d^{{}}_0 &=\:&  9.9983952  &\times10^{2}, &
 d^{{}}_3 &=\:& - 4.6170461 &\times10^{-5}, \nn \\
 d^{{}}_1 &=\:& 1.6945176 &\times10^{1},&
 d^{{}}_4 &=\:&  1.0556302  &\times10^{-7}, \nn \\
 d^{{}}_2 &=\:& -7.9870401 &\times10^{-3},  \quad &
 d^{{}}_5 &=\:&  -2.8054253  &\times10^{-10},\nn
 \end{alignat}
and where the apparent specific volume $V^{{}}_\mathrm{app}(S,T)$ of NaCl is given by
 \beq{Vapp}
 V^{{}}_\mathrm{app}(S,T) =
 \frac{S + e^{{}}_2 + e^{{}}_3 \frac{T}{1~\SICel}}{
 (e^{{}}_0 S + e^{{}}_1)\:\exp\!\bigg[
 e^{{}}_5 \Big(\frac{T}{1~\SICel} + e^{{}}_4\Big)^2
 \bigg]}\:\SIm^3\:\SIkg^{-1},
 \eeq
with the coefficients $e^{{}}_j$ being
 \begin{alignat}{4}
 e^{{}}_0 &=\:& -4.330  &\times10^{-2}, &
 e^{{}}_3 &=\:& 1.4624 &\times10^{-2}, \nn \\
 e^{{}}_1 &=\:& 6.471 &\times10^{-2}, \quad&
 e^{{}}_4 &=\:&  3.3156  &\times10^{3}, \nn \\
 e^{{}}_2 &=\:& 1.01660 &\times10^{0},  \quad &
 e^{{}}_5 &=\:& 1.0000  &\times10^{-6}.\nn
 \end{alignat}

The viscosity $\etaScs(S,T)$ of NaCl/water solutions is given by Lalibert\'e \cite{Laliberte2007} as
 \beq{etaScs}
 \etaScs(S,T) = \big[\etaWa(T)\big]^{(1-S)}
 \:\big[\etaNaCl(S,T)\big]^S,
 \eeq
where the viscosities $\etaWa(T)$ and $\etaNaCl(S,T)$ of water and liquid NaCl, respectively, are
 \bal
 \eqlab{etaWa}
 \etaWa(T) & = \frac{\big[\frac{T}{1~\SICel} + 246 \big]\:\SImPas}{
 \big[0.05594\frac{T}{1~\SICel} + 5.2842\big]\frac{T}{1~\SICel} + 137.37},\\
 \eqlab{etaNaCl}
 \etaNaCl(S,T) &= \exp\Bigg[\frac{h^{{}}_1 S^{h_2} + h^{{}}_3}{
 \big[1+h^{{}}_4\frac{T}{1~\SICel}\big]\:\big[
 1 + h^{{}}_5 S^{h_6}\big]} \Bigg] \SImPas,
 \eal
with the coefficients $h^{{}}_j$ being
 \begin{alignat}{4}
 h^{{}}_1 &=\:& 1.6220 &\times10^{1}, &
 h^{{}}_4 &=\:& 7.4691 &\times10^{-3}, \nn \\
 h^{{}}_2 &=\:& 1.3229 &\times10^{0}, \quad&
 h^{{}}_5 &=\:& 3.0780 &\times10^{1}, \nn \\
 h^{{}}_3 &=\:& 1.4849 &\times10^{0},  \quad &
 h^{{}}_6 &=\:& 2.0583 &\times10^{0}.\nn
 \end{alignat}
%

%
%

%

\end{document}